# Le travail coopératif comme vecteur d'évolution de nos systèmes d'information

*Patrick Nourrissier*

NETLOR
14, Bld du 2&ième RA  5400 Nancy
France

patrick@netlor.fr

*Sahbi SIDHOM*

MCF. & Chercheur
LORIA–Nancy2 & Nancy Université
France

Sahbi.Sidhom@loria.fr

**Résumé :**

Cet article porte sur la présentation de l'outil coopératif Dims, une plate-forme qui permet de gérer le quotidien d'un acteur, en lui apportant souplesse et rapidité dans l'organisation de ses tâches. L'intérêt principal réside dans la possibilité d'organiser le système d'information selon une logique réseau, réparti sur plusieurs sites physiques, en y intégrant un protocole spécifique rendant possible la communication inter-serveurs. Ce protocole, basé sur le standard Jabber, répond à des besoins de collaboration simples et quotidiens, en permettant la distribution de recherche et d'accès aux ressources de ce réseau étendu. L'objectif technologique concerne l'évolution des architectures des systèmes d'information, où l'applicatif pourra être demain un ensemble complet de services et outils distribués sur un réseau de sites distants. L'enjeu pour l'utilisateur est la perception d'un rôle collectif pour chaque individu qui coopère, dans le respect des règles de sécurité et sans aucune limite technique d'accès à l'information.

**Mot clé :** travail coopératif, système d'information (SI) distribué, partage d'information, superposition des contextes de l'information, collaboration laboratoire/entreprise.

# Le travail coopératif comme vecteur d'évolution de nos systèmes d'information

**1. Introduction**

Nombreux sont les termes qui caractérisent aujourd'hui ce besoin de comprendre, gérer et capitaliser nos connaissances, acquises et entretenues tout au long de notre parcours professionnel. En tant qu'acteur et consommateur de notre environnement économique, nous exprimons ce souhait à travers nos usages quotidiens de l'information, qu'ils soient personnels ou professionnels.

Notre usage personnel de l'information est rythmé par le progrès technique de cette dernière décennie, grâce notamment à l'ouverture des réseaux de communication et leurs capacités techniques à rendre notre système informatique communicant. La progression massive des personnes ayant un accès à Internet est d'ailleurs un bon indicateur de cette progression technologique. Dans la pratique, cette possibilité se focalise à quelques composants logiciels, simples et efficaces, tels que la messagerie, la navigation sur Internet ou encore la messagerie instantanée. Tous ces outils exploitent pleinement ces nouveaux moyens de communication et rendent plus accessibles les usages de l'informatique.

Dans un contexte professionnel, les usages diffèrent. Ils dépendent fortement de l'environnement informatique installé. L'évolution de nos SI suit un rythme beaucoup plus lent, nécessitant bien souvent le recours à des entreprises extérieures pour innover et migrer le SI existant [1]. A la perte possible de données s'ajoute celle du savoir intrinsèque créé autour de l'application existante. Ce savoir détenu par les acteurs (employés de l'entreprise) nécessite pour le préserver un accompagnement au changement [2]. Malheureusement, ce savoir implicite est peu connu, peu décrit et rarement intégré.

Un exemple simple de ce constat est la carte de visite. Que vous soyez en conférence, sur un salon ou encore en rendez-vous clientèle, les cartes collectées finissent généralement dans un tas intemporel de contacts et événements, entretenus par quelques recherches souvent infructueuses. Cette gestion non structurée dilue notre connaissance de ces événements et contextes passés. Pour palier à ce problème, la tendance se situe dans le partage d'information, à l'aide de partenaires potentiels sur la toile ou par le biais de réseaux spécialisés dans le recoupement d'informations diluées.

**2. Problématique**

En nous appuyant justement sur l'usage de ce type d'échange tel que les réseaux sociaux, où chaque individu veut satisfaire des besoins personnels et collectifs, nous pourrions décrire ce formalisme d'échanges et de coopération comme une forme d'altruisme réciproque. Chaque acteur possède un intérêt à capitaliser et partager son information, qu'elle soit ou non décrite dans le contexte de sa création. L'objectif est surtout d'affiner ses connaissances [4] et de tester leurs fiabilités par recoupement [3]. Il reste à savoir ce que l'on souhaite donner au regard de ce que l'on obtient. La prudence est de rigueur, et ce partage s'effectue principalement par étape, afin d'éviter les pièges d'un déséquilibre trop important.

Cette problématique se rencontre quotidiennement en entreprise, même si les raisons diffèrent. Au sein d'une même organisation, les acteurs ont un intérêt commun à partager, sans forcément disposer de moyens techniques suffisants. Dans de nombreuses petites et moyennes entreprise, les supports physiques de stockage résident dans un serveur de fichiers unique, accessible sur le réseau, et contenant l'ensemble des archives de la société. Dés lors, on peut s'interroger sur le devenir de toutes ces informations collectées sur des supports de plus en plus mobiles, tels que les clés USB, les ordinateurs et téléphones portables. Chacune de ces unités possède les capacités à créer de l'information qui, si elle n'est pas intégrée dans le SI, ne fera pas partie du patrimoine informationnel. Nos usages nous poussent à simplifier notre gestion quotidienne de l'information au détriment de sa structuration. Cette tendance très forte impacte notre organisation de travail. Aujourd'hui, nous devons concevoir des systèmes d'information tenant compte de ces supports et usages.

L'enjeu est donc d'aider les utilisateurs dans la gestion quotidienne de leurs informations. Nos travaux tentent d'améliorer la gestion individuelle et collective de celles-ci, en exploitant au mieux les nouvelles technologies. L'objectif à moyen terme est d'établir des méthodes simples de coopération qui guideraient nos actes et permettraient la superposition de nos contextes d'usages.

Pour répondre à ces enjeux et problématiques, la première barrière à lever est l'harmonisation des accès et supports de l'information. Nous nous sommes basés sur la logique de la messagerie instantanée. Ce principe permet à chaque individu d'être prévenu de la disponibilité de l'autre, d'échanger de l'information en mode synchronisé ou non, et de distribuer une demande à plusieurs serveurs formant ainsi un réseau étendu. En partant du principe que chaque acteur du système possède un accès à ce réseau, nous organisons notre architecture système sur ce principe d'inter-connexion, où l'enjeu n'est pas tant de mettre en réseau ces acteurs que d'imaginer la disponibilité permanente de l'ensemble des ressources.

Pour exemple, on peut imaginer qu'une personne en déplacement souhaite récupérer la dernière version d'un document pour le modifier, la première question qu'il pourrait se poser est : où est ce document et où se trouve la dernière version ?

L'idée est d'exploiter ce réseau pour identifier la ressource modifiée, son auteur et sa localisation. La personne peut donc télécharger la dernière version depuis le lieu d'origine, le modifier et surtout l'envoyer en un clic à son lieu de stockage d'origine après modification. Nous pouvons facilement suivre la vie de ce document et surtout éviter la création de richesse locale en dehors du SI. Par extension, la cartographie de l'information en serait grandement simplifiée, permettant à chaque individu de retrouver plus facilement une information dont le lieu est clairement identifié. De même, des services métiers spécifiques peuvent être dédiés à certaines ressources, rendant l'exploitation plus aisée.

**3. Outil proposé**

Pour permettre toutes ces interactions, une plateforme a été développée, appelée Dims portal. Cet outil de travail coopératif, vous aide à gérer votre quotidien, en vous apportant souplesse et rapidité dans l'organisation de vos tâches quotidiennes. Ce framework de développement est fondé sur des technologies ouvertes (PHP/JAVA, Mysql et Linux).

Cet outil s'adapte au modèle organisationnel de votre structure, en proposant la création d'espaces de travail (cf. Fig. 1, 2). Chaque espace peut créer ses propres briques métiers, appelées modules, ou utiliser des modules partagés. La collaboration s'effectue à l'aide de ces espaces de travail où chaque information bénéficie de fonctions simples de collaboration.

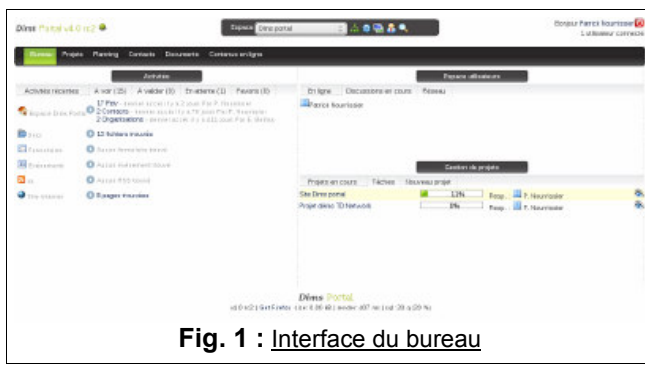
**Fig. 1 :** Interface du bureau

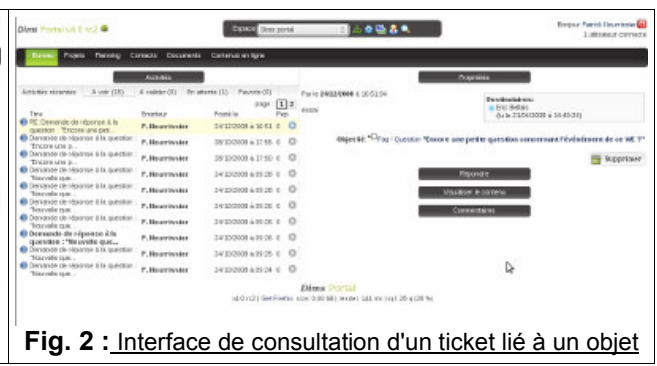
**Fig. 2 :** Interface de consultation d'un ticket lié à un objet

L'objectif principal est d'améliorer et de minimiser les échanges d'information (éviter l'asphyxie des messageries), tout en mettant en avant le contexte dans lequel les annotations ont été produites. Une différenciation entre l'action à produire et l'information à voir est mise en évidence, afin de gagner en efficacité et surtout favoriser la collaboration entre les acteurs (cf. Fig. 3, 4).

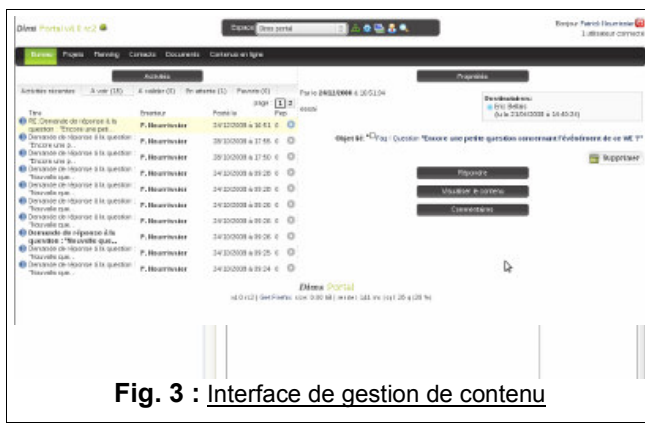
**Fig. 3 :** Interface de gestion de contenu

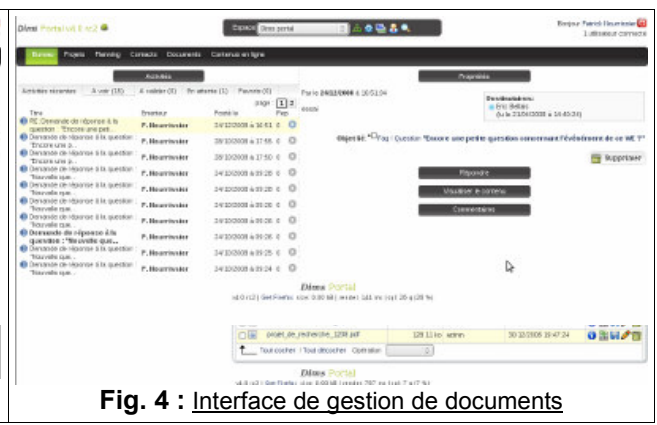
**Fig. 4 :** Interface de gestion de documents

Quelques éléments clés ont été développés, tels qu'un planning et une gestion de projets, dans l'esprit de la coopération. Un moteur de recherche évolué, basé sur l'indexation de l'ensemble des contenus et contextes, permet l'interprétation de recherche à l'aide de parenthèse et d'opérateurs booléens. Il propose également la possibilité de mettre en surveillance une demande spécifique afin d'analyser les nouvelles informations saisies (cf. Fig. 5, 6).

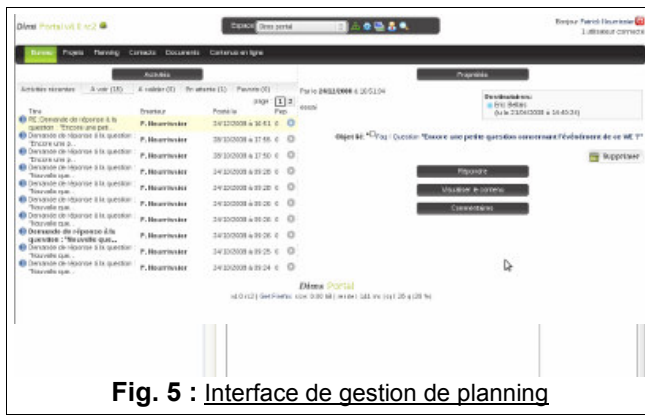 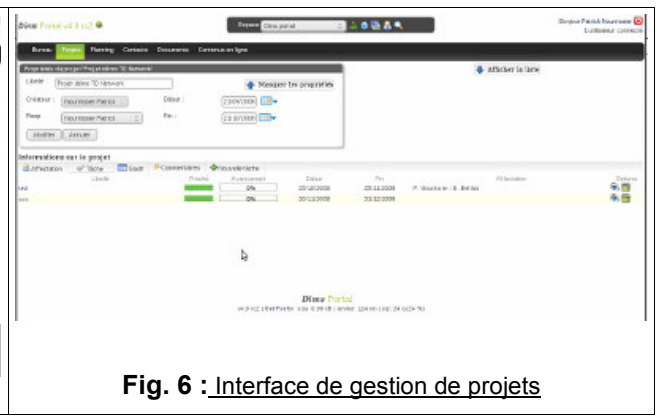

**Fig. 5 :** Interface de gestion de planning    **Fig. 6 :** Interface de gestion de projets

**4. Travaux de recherches en cours**

Outres ces fonctionnalités, l'intérêt de cette plateforme est sa capacité à dialoguer entre plusieurs serveurs, rendant ces fonctionnalités de recherche, d'accès et d'échanges possibles sur le réseau constitué. Cet intérêt répond à un besoin croissant d'acteurs, ayant des comptes multiples sur des dispositifs qui doivent être séparés. Pour ce faire, nous avons fait le choix du protocole Jabber, qui offre toutes les fonctionnalités nécessaires à ces échanges. Le schéma suivant (cf. Fig. 7) présente une organisation possible de plusieurs dispositifs Dims avec des accès différents (navigateurs Internet ou client léger de type msn messenger).

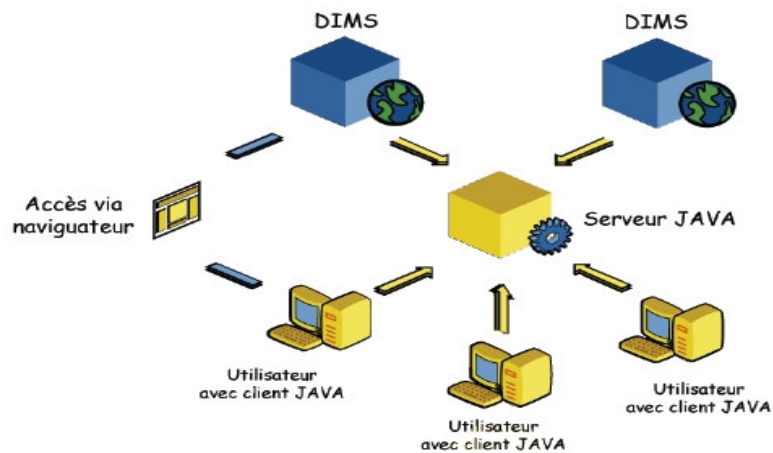

**Fig. 7 :** Organisation possible de plusieurs dispositifs Dims

**5. Bibliographie**

[1] David A., Thiery O., « Prise en compte du profil de l'utilisateur dans un système d'information stratégique », VSST'2001.

[2] David A., Thiery O., Nourrissier P., « De l'élaboration d'un site web à l'extraction de données », EGC'2002.

[3] Péguiron, Frédéric et Thiery, Odile. « Modéliser l'acteur dans le système d'information stratégique d'une université. In Veille Stratégique Scientifique et Technologique » *VSST'2004*. 2004. 5p.

[4] ZACKLAD, M., « Ingénierie des connaissances appliquée aux Systèmes d'Information pour la coopération et la gestion des connaissance », Habilitation à diriger des recherches, Université Paris 6, 2000.